# Interconnectedness in Education Systems

Cristian Candia[1,2,*] (0000-0002-9654-543X), Javier Pulgar[3] (0000-0003-0325-7306), Flávio L. Pinheiro[4] (0000-0002-0561-9641)

[1]Data Science Institute, Facultad de Ingeniería, Universidad del Desarrollo, Las Condes, 7610658, Chile
[2]Northwestern Institute on Complex Systems (NICO), Northwestern University, Evanston, IL 60208
[3]Physics Department, Universidad del Bio Bio, Concepcion, Chile
[4] NOVA IMS -- Universidade Nova de Lisboa, Lisboa, Portugal
* Corresponding Author: crcandiav@gmail.com

Underlying complex systems, there is a rich web of interconnected components that determine the relational properties of the system. Yet, traditional methods used in education sciences often disregard the underlying complexity of the educational system and, consequently, its emergence phenomena. Here, we argue that an interconnected vision of educational systems -- from classrooms to an organizational level -- is key to improving learning, social integration, well-being, and decision making, all fundamental aspects of the educational experience. Understanding the education system as an interconnected network of people, degree programs, and institutions requires methods and concepts from computational social sciences. Thus, we can leverage institutional records and (quasi) experimental designs to elicit the relational maps of key players in education and derive their implications in their functioning at all scales. In different settings, from elementary classrooms to higher education programs, we show how mapping the network relationships between entities can lead to the inference of novel insights about education systems and the development of solutions with societal implications.

## Introduction

"The education system is one of the most complex and challenging systems for research" (Lemke and Sabelli 2008) (p. 128). Recent trends and crises around the globe —technological changes, growing inequality, and pandemics— have produced new challenges to schools, universities and, in general, to the entire educational system (OECD 2020). The growing demand for accessing education across the world forced society to expand educational opportunities beyond the exclusive elites to a vast majority. Therefore, nation-wide systems and communities face issues regarding students' learning processes (performance and creativity), social integration (cooperation, friendship, and bullying), and educational outcomes (dropouts-retention, gender inequality, and schooling choice), and the effectiveness of pedagogical practices (decision-making) throughout all education levels. Such challenges are the consequence of the existing relational processes among people, such as the simple act of sharing information with family, friends, and colleagues. Here, traditional views of educational research fail to take advantage of emergent properties of human dynamics in classrooms, schools, and universities. Naturally, such inability to link relational processes from individual behaviors poses a challenge to our ability to understand human dynamics using standard tools not equipped to deal with the complex nature of education. Therefore, a new framework capable of integrating the different relational aspects of education is necessary to formulate and assess innovative practices and interventions.

An essential feature of complex systems is that collective behavior cannot be inferred from the sum of their constituents. Instead, it leads to the emergence of new, and often unexpected, non-linear and unpredictable behaviors (Baas and Emmeche 1997; Bar-Yam 2002). The concept of emergence is closely related to the concept of criticality and self-organization in Physics (Bak 2013), which is in turn linked to chaos theory (Lansing 2003). In social sciences, emergence phenomena is often associated with the concepts of human capital (Arena and Uhl-Bien 2016), innovation (Lane et al. 2009), and economies of scale (Arthur et al. 1994; Beinhocker 2006). For instance, increasing agglomeration of individuals provides increasing returns in terms of innovation outputs such as patents and scientific publications (Balland et al. 2020). Interestingly, underlying complex systems is often, if not always, a complex network structure that defines how its elements interact and affect the functional properties of the overall system dynamics (Barabási 2002). For instance, social interactions and the ensuing diffusion of ideas, opinions, and behaviors.

In education, the social network formed by students and teachers defines the backbone of education as a complex system. Suchlike network structure plays a critical role in defining the effectiveness of interventions (Valente 2012) and in shaping the functioning of the system (Barabási 2002) (e.g., social, educational, economical, ecological, etc.). The concept of the educational system as being supported by a social network composed by its main actors implies that new methods are needed to take into account the natural complexity of the processes that unfold in the educational setting. Moving from traditional approaches which ignore the existence of such networks, scholars suggest combining conceptual perspectives about complex systems with computer modeling capabilities (i.e., simulation and big data analytics) to inform teachers and educators about proposed interventions and their potential impact (Jacobson, Levin, and Kapur 2019; Mason 2008). For instance, measuring students' disposition to cooperate in the classroom and its impact on performance is challenging, and requires multidimensional instruments to reveal various forms of social relationships. Such efforts to capture social networks have their origins in Jacob Moreno's sociograms (Moreno 1934), obtained to map who students and their friends like to share time with (Coie, Dodge, and Kupersmidt 1990; Mouw 2006; Neal 2007; McCormick and Cappella 2015; Ivaniushina and Alexandrov 2018). More recent evidence on networks in education had characterized students' social structures in various learning contexts (Traxler et al. 2020; Commeford, Brewe, and Traxler 2021; Javier Pulgar, Rios, and Candia 2019), and linked students' social networks with their academic performance (Putnik et al. 2016; J. Bruun and Brewer 2013; Blansky et al. 2013; J. Pulgar, Candia, and Leonardi 2020; Javier Pulgar et al. 2021; Javier Pulgar, Rios, and Candia 2019; Candia et al. 2019). Besides, the use of network methods in students' applications to higher education has fostered new insights into the structures that govern accessibility and opportunities in post-secondary education (Candia, Encarnação, and Pinheiro 2019).

We argue that network analysis of educational systems constitutes an important contribution to the study of a wide variety of educational processes. The growing and ubiquitous technology penetration and access to information on student populations and schools' performance becomes an opportunity for designing effective responses and adaptive capabilities to face societal and environmental changes. Thus, with the increasing availability of data and our ability to process and analyze it, the use of network science methods is a fundamental framework to expand our current understanding of educational systems from classroom dynamics to higher-level governance and policy related topics. In the following sections we begin with an overview of past contributions of network analysis to education, and we explore two avenues that, in our opinion, exemplify promising fields for future research: social relationships in education for human development; and the interconnected nature of higher education systems for individual and system-wide decision-making.

## Networks in the Classroom and Students' Performance

The attention to social networks in education has fostered educators and scholars to dive into a myriad of perspectives to expand our understanding of the educational experience (Grunspan, Wiggins, and Goodreau 2014). If traditional learning frameworks and methodologies focused on individual attributes to explain performance, attention to networks has enabled a unique access to the complex social dynamics in the classroom (Commeford, Brewe, and Traxler 2021; Traxler et al. 2020; J. Pulgar 2021), in line with the characteristics of the socio-cultural theory of human development (Vygotsky 1978), social cognitive theory (Bandura 2001), and the principles of social capital (Lee 2010). One of the main drivers for using network analysis in education is to understand how the interplay between social ties and performance, such as whether centrality and the structural properties of the classroom and school are related with success (J. Pulgar, Candia, and Leonardi 2020; Putnik et al. 2016; J. Bruun and Brewer 2013; Calvó-Armengol, Patacchini, and Zenou 2009a) and university retention (Zwolak, Zwolak, and Brewe 2018). Moreover, studies have centered around the effects of peer influence and friendship on achievement (Eckles and Stradley 2012; DeLay et al. 2016; Stadtfeld et al. 2019c), as well as the emergence and change in the number and nature of social ties (Brewe, Kramer, and Sawtelle 2012; Eagle, Pentland, and Lazer 2009; Smirnov, and Thurner 2017), or the social structures fostered by certain teaching methodologies (Commeford, Brewe, and Traxler 2021; Traxler et al. 2020).

As mentioned, peer effects are critical for both social learning (Henrich 2015; Gil-White and Henrich 2001; Pentland 2015; Johnson and Johnson 1987; Roger and Johnson 1994), and academic outcomes (Kassarnig et al. 2018; Blansky et al. 2013; Sacerdote 2011; P. Davies 2018; Biancani and McFarland 2013). Social learning becomes a natural form of pedagogy between learners, as cultural knowledge is shared via communication, imitation (Csibra and Gergely 2011), or social contagion, as individuals who enjoy high exposure to information, behaviors, attitudes and beliefs are more likely to adopt them in comparison with more isolated peers (Bandura 1986; Freedman and D. 1978; Burgess et al. 2018; Benson and Gresham 2007). This form of pedagogy requires both the "student-role subject" along with the motivation to share knowledge typically observed at the teacher level (Csibra and Gergely 2011), and thus qualifies as a basic form of collaboration captured in students' social relationships.

Past works have found correlational evidence between students' network positions and academic performance at different levels of education (Baldwin, Bedell, and Johnson 1997; Caprara et al. 2000; Jesper Bruun and Brewe 2013; Gašević, Zouaq, and Janzen 2013; Blansky et al. 2013; Ivaniushina and Alexandrov 2018; Stadtfeld et al. 2019b; Kassarnig et al. 2018; Berthelon et al. 2019). For instance, in online university programs, achievement has shown to positively correlate to measures of social capital (Gašević, Zouaq, and Janzen 2013), along with interchanging information through online and offline communication platforms (Kassarnig et al. 2018). Yet, the relationships between students' centrality and their performance is not quite clear. The assumption that enjoying a higher number of social ties in the classroom results in academic success is challenged by the nature of the task and learning environment (J. Pulgar, Candia, and Leonardi 2020). As students' outcomes mirror the affordances of structural holes for creative combinations (Burt 2004), or highly constrained networks (Rhee and Leonardi 2018) depending on whether the activities are open-ended or close-ended, respectively. Figure 1 depicts the collaboration network of three university classrooms, with information on grades on traditional physics problems (color nodes) and degree centrality (node size). As shown, bigger and central nodes are not uniquely the ones that hold good performance, particularly in the traditional section (for more details see (J. Pulgar, Candia, and Leonardi 2020)). Additional evidence has found that reciprocal collaborative ties bring higher performance returns in elementary education using a standard social dilemma from game-theory, while unidirectional ties are negatively related to grades (Candia et al. 2019).

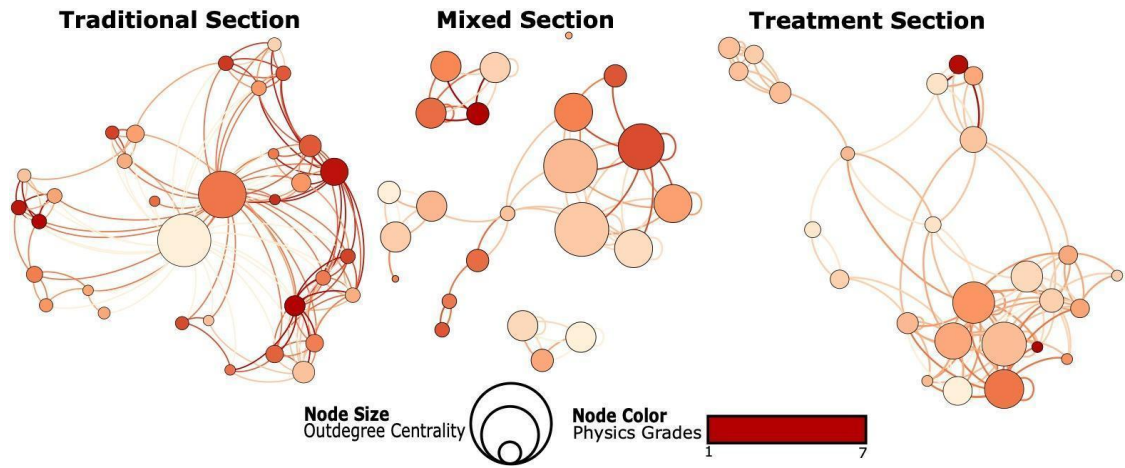

*Figure 1: Classroom networks for the three analyzed sections: traditional, mixed, and treatment. Node color represents physics grades (dark red being the highest), and the node size represents the out-degree centrality, i.e., the number of times that a student seeks information in the classroom. Note. Reprinted from "Social networks and academic performance in physics: Undergraduate cooperation enhances ill-structured problem elaboration and inhibits well-structured problem solving", by Pulgar et al., 2020.*

Moreover, students' networks and their performance are sensitive to socially constructed expectations, as motivations for social status and popularity relate to learning achievements and performance outcomes (e.g., good grades) (Vignery and Laurier 2020; Stadtfeld et al. 2019a; Rizzuto, LeDoux, and Hatala 2009), respectively, but through different motivations (Jones and Cooke 2021). Learning achievements are linked to the adoption of optimal behavioral interactions, whilst a focus on getting good grades is accompanied by competitive comparison and self-improvement. Interestingly, when accounting for classroom-level motivation, friendship has shown to influence performance only in classrooms that seek to develop mastery rather than high scores (Laninga-Wijnen et al. 2018). Thus, in the pursuit of human development through social interactions, it is worth attending to the socially constructed set of norms and behavioral expectations constructed within the classroom or school (J. Pulgar 2021).

Social justice and inclusion in schools present an interesting case of looking at education from the lens of networks. Within the literature it is possible to find evidence of the overwhelming agreement regarding the benefits of group processes and collaboration for individual growth and development on both students (Zandena, Meijera, and Beghetto 2020; D. Davies et al. 2013) and educators (López Solé, Civís Zaragoza, and Díaz-Gibson 2018, 2018; Moleenaar and Sleegers 2010). Yet, and as presented earlier, the effectiveness of such collaborative learning settings may be optimized by utilizing network analytics, for instance, in the identification of academic and social hierarchies within a system. Such information can then be used to design strategic mechanisms for those left aside from the social ladder, and thus marginalized from the benefits of social inclusion. Isolated individuals are more at risk of academic failure (Zwolak, Zwolak, and Brewe 2018), or in the case of educators, have limited access to professional resources and motivation for engaging in the adoption of school reforms and innovations (Moleenaar and Sleegers 2010). Attention to these individuals and the overall social structure within schools is critical at various levels of education: in and outside the classroom; and within communities of educators and administrators. Consequently, utilizing networks to map out relationships and design strategic mechanisms for inclusion becomes a key affordance to optimize learners' and educators' social capital in the spirit of strengthening their communities of practice.

On these scenarios, student networks can be measured by administering either quantitative (Traxler et al. 2020; Javier Pulgar et al. 2021), qualitative (Borgatti, Everett, and Johnson 2013; Sundstrom et al. 2022), or mixed-method protocols (Froehlich 2020). There are many strategies for network data collection. Yet, it is particularly relevant to focus on entire networks at the classroom level, as the symbolic space that constitutes the classroom provides the well-bounded limits for the social system. The most frequent strategy is network surveys where participants could either write down the names of their friends or working peers (i.e., sociometric instrument) or by selecting their ties from the roster of individuals that formed the network under study (i.e., census) (Carolan 2014). Of course, each of these strategies cannot capture the qualitative nuances embedded in different social ties. Therefore, researchers have opted for designing qualitative protocols, such as interviews (Borgatti, Everett, and Johnson 2013), observations (Sundstrom et al. 2022), or mixed methods strategies that combine the shortcomings of surveys with the costs of coding processes (Froehlich 2020).

Interventions targeted on social and emotional skills have yielded positive outcomes, as individuals develop mechanisms to form friendships and overcome the potential bias of salient attributes (e.g., gender and race), thus, creating a more optimal learning culture (DeLay et al. 2016). Network based interventions, such as random group assignment among first year university students fosters initial friendship ties (Boda et al. 2020), which are likely to evolve and transform into study relationships (Stadtfeld et al. 2019c) as academic hierarchies are revealed, and students seek out to form ties with peers that have similar performance (Smirnov, and Thurner 2017).

As such, the potential to quantify human interactions using network analysis has been critical to comprehend the extent to which learning and development are encouraged by one's social integration, but also, stimulates the inclusion of social dimensions in the assessment of data-driven educational interventions and policies (Goertzen, Brewe, and Kramer 2013).

## Networks in Elementary Education

Long-term returns to education depend mostly on early learning outcomes (Claessens and Engel 2013; Dickens, Sawhill, and Tebbs 2006; Restuccia and Urrutia 2004) that motivate the study of elementary school populations. Traditionally, relational studies in older populations are conducted through surveys. However, for elementary school students, survey-based social network mapping may not be sufficient because survey-based instruments have issues, such as the social desirability bias (Van de Mortel et al. 2008) (over-reporting of socially desirable behavior); cognitive barriers (difficult to establish that subjects fully understand the questions) (Borgers, Leeuw, and Hox 2000); and lack of engagement associated with the implementation of the instruments (length or unfriendliness of instruments generate poor answers) (Banister and Booth 2005; Barker and Weller 2003; Fleer and Ridgway 2014; Kyritsi 2019; Leeuw 2011). An additional challenge is to study populations as young as primary school children. However, experimental game theory can help to overcome all these issues.

Social relationships among elementary school students can be mapped using a video game based on game theory. This methodological approach facilitates the behavioral mapping of cooperative relationships by settling elementary school students in a more familiar and ecological interactive environment. The advantage of using experimental game theory to map the social network is twofold: first, it allows to directly capture cooperative relationships among students; and second, the interactivity of the game, in which different actions lead to different payoffs, partially mitigates the aforesaid issues of survey-based instruments (Colin F. Camerer and Hogarth 1999). The experimental approach recreates social exchange and forces an action with (virtual) consequences towards students' counterparts, mitigating social desirability biases prevalent in declarative responses of survey-based instruments.

## Experimental Game Theory Approach

In the experimental game theory setup for measuring cooperation and reciprocity, the lab is brought to the field using a friendly tablet computer interface (Fig. 2A) boosting engagement and mitigating cognitive barriers (Colin F. Camerer 2003) associated with the young age of elementary school samples. Thus, gamification of didactic materials and use of a friendly interface implemented using tablet computers is preferred to standard experimental interfaces such as z-tree.

The video game implements a modified Prisoner's Dilemma where students played a non-anonymous and simultaneous dyadic game once with each of their classmates (Fig. 2B). Under standard game-theoretic experimental protocols, which involve anonymous interaction, the elicited networks in the lab emerge from scratch, mainly through assortative interaction between anonymous players (see, for instance, (Goeree, Riedl, and Ule 2009)). However, in order to capture the pre-existing relationships, we have to depart from the standard protocol and consider non-anonymous interactions (Wang et al. 2017; Candia et al. 2019), i.e. in our social dilemma, students are aware of the identity of the partner that has been matched to them each round. Thereby, their choices when playing against each other are not only the result of intrinsic prosocial dispositions (or their absence), but also the result of their history and the perceptions they have about each other.

Students play the dyadic social dilemma among all of the possible dyads in a class group. Students are endowed with 10 tokens per round, that they can share or keep (cooperate or defect). Each student plays with a different classmate and, simultaneously, they have to decide how many tokens to send to their peer. A multiplicative factor is added on donations to allow for potential efficiency gains; therefore, there is a tradeoff between reaping the benefits of cooperation by sending all tokens and the risk of exploitation by non-reciprocators. Hence, the dyad maximizes the total number of tokens earned when everyone cooperates, but students maximize their tokens when they defect and everyone else cooperates (Fig. 2B).

In summary, students have to drag and drop tokens to their peers, and the game follows as:

- In each dyadic interaction both students are endowed with ten tokens.
- Each student decides how many tokens to give to the other students.
- Each sent token doubles, creating an incentive to cooperate, only if you expect cooperation to be reciprocated.
- Finally, each student plays with each of their classmates one round at a time, repeating all the previous steps.

Collecting all the cooperative interactions allows us to map students' networks. For instance, Figure 2C shows a classroom network in which nodes represent students and directed links indicate fully cooperative interactions (a student giving ten tokens to another student). The experimental approach can be complemented with other types of information sources, such as peer-nominations surveys, that allow for the explicit declaration of different categories of relationships. For instance, identifying friends and also obtaining information about the set of attributes that a student has as declared by her/his school mates, and thus, identifying the most popular/aggressive students according to their peers. Institutional records are also a valuable information source which encompasses all the data regarding academic trajectories such as, class attendance, class group belonging, and grades.

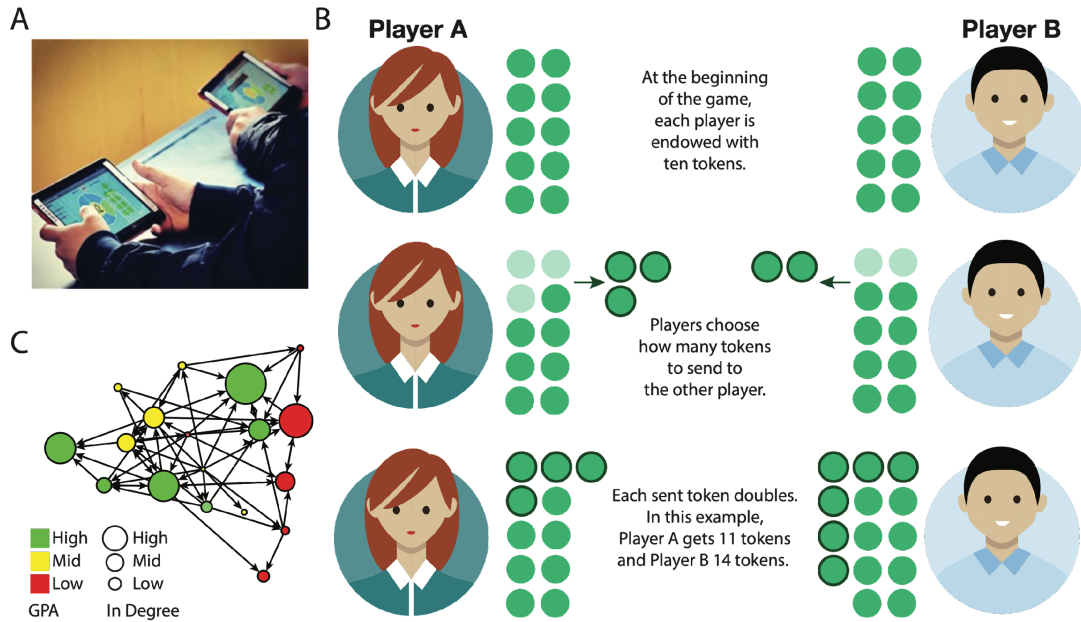

*Figure 2: Experimental game. A) Video game interface. B) Social dilemma: i) Both players are endowed with 10 (ten) tokens. ii) Simultaneously, player A decides to send 3 (three) tokens, and player B decides to send 2 (two) tokens. iii) Sent tokens are doubled. iv) Player A receives 4 (four) tokens and player B receives 6 (six) tokens. C) Students' network edges represent fully cooperative interactions. Note. Reprinted from "Strategic reciprocity improves academic performance in public elementary school children", by Candia et al., 2019.*

Hence, it is possible to explore questions involving different sources of relational data. If the outcome of the experimental game can predict declarations of mutual friendships in the peer nomination instrument, we can understand how gradual measures of the tie's strength and the level of symmetry of such friendship correlate with such declarations. For example, the relationship between donations towards a given "recipient" and her/his position on the friendship network (Goeree, Riedl, and Ule 2009). The information collected about the set of attributes relevant in the context of social life (popularity, aggressiveness) might also help us to improve our understanding of the relationship between experiment-based mapping of the topological properties of the student's position in the network and her/his academic performance and other educational outcomes.

Evidence on the the external validity of game-theoretic experiments (Karlan 2005; Fehr and Leibbrandt 2011; Gelcich et al. 2013; Hopfensitz and Miquel-Florensa 2017; Algan et al. 2013) shows that real-life cooperation is correlated with cooperation under laboratory conditions. Thus, we expect that these cooperative dispositions at the individual level will be revealed by the behavior implemented in the games. The main difference between standard experiments and our protocol comes from the non-anonymous feature of laboratory interactions. Thereby, we expect to capture not only the individual's intrinsic disposition to cooperate, but also their dispositions to cooperate within the context of a particular relationship and their history together (Wang et al. 2017). For instance, by using this approach in a population of 1,000 elementary school students it was possible to show that high GPA students cooperate more strategically with their peers, sending more cooperation to their friends, compared to low-GPA students, which translates into more cooperative returns (Candia et al. 2019). Also, students who participated in more reciprocal relationships improved their GPAs faster (Candia et al. 2019). Finally, social hierarchies emerge and peers cooperate following a simple rule, the higher the difference in social hierarchies the higher the difference in sent tokens unless friendship exists between peers. In other words, low-ranked students cooperate more with high-ranked students if they are not declared as friends. If friendship exists, both students cooperate the same regardless of their social rank difference (in-review paper).

## The Interconnectedness of the Higher Education System

Moving away from classroom applications to organizational aspects of higher education, network analysis offers a suitable framework to map the interdependence between the different elements —students, educators, schools, majors, degree programs— that constitute education systems and, consequently, unfold their organizational principles. In that sense, networks provide a natural framework to formalize relationships between elements that would otherwise be unfeasible or unnatural to explore. For instance, what is the distance between degree programs in Physics and Sociology? How related is Physics with Chemistry? Certainly, one can compare curricular content or the historical roots and branches of disciplines to obtain an answer to these questions. However, the assumption that similarity and relatedness between degree programs is solely defined by their topics ignores other less obvious properties that might make them more or less related in the eyes of the actors that interact with education systems. Naturally, factors such as employability, individual and group preferences, market needs, costs, trends, popularity, and spatial distribution of institutional offer play a role in how students and, certainly, educators and policymakers should relate different degree programs. More importantly, what question are we trying to answer when measuring distances/relatedness? Are we looking to extend the curricular offer of our school, to recommend viable choices to applicants/students, or predict the impact of an intervention such as changing the number of open positions (e.g., numerus clausus)?

In that sense, Baker (2018) proposes mapping the proximity between majors offered in US Community Colleges as a means to identify groups of similarly preferred majors, thus aiding colleges in the design of guided learning pathways and meta-majors (i.e., groups/buckets of majors and courses). This approach would contrast with the traditional open choice format offered in US universities, where students can freely choose in which courses to enroll and in which major to graduate. As pointed out, such free choice, although flexible and able to accommodate the contextual education needs of each student, can also contribute to the high dropout rates among college students by its inherently lack of pedagogical structure. In that context, Baker (2018) gathered data through questionnaires to understand how students cluster together different majors in a College. The resulting structure is explored using network analysis. Results reveal a modular structure that could help educators and schools to better design meta-majors in light of their perceived similarity.

Moreover, Baker (2018) suggests that perception mismatches between students and educators or decision-makers need to be taken into consideration when developing new policies. While educators often rely on rigid taxonomies and organizational views designed by panels of experts to support their policies and derive comparative statistics, such descriptions do not have to necessarily match the true perception of the actors involved (i.e., applicants, students, educators). Hence, computational social scientists should look for richer data-driven approaches to quantify the perceptions of the different actors in light of what is known about individual's choice and activity (Chaturapruek et al. 2021). For instance, through data it is possible to proxy the revealed preferences, attitudes, and choices of individuals and offer a complementary view of the educational system. Such data often is already stored in existing schools and institutional databases. For instance, data on students' participation in discussion forums, their enrollment choices and academic performance, or their participation in group activities are examples of datasets that schools have on their possession but rarely use. These datasets offer valuable proxies to infer interactions between students but also to map information flow and peer-influence dynamics, which can be used to assess how student's performance is affected by their interactions, participation in activities, but also on underlying factors that explain their interplay with educational systems As such, by embracing computational and network methods to study existing data, computational social scientists can derive a description that is closer to individual perceptions, with the advantage of being more flexible and capable of temporal variations in individuals choices, preferences, and attitudes towards the educational system.

In that sense, Candia, Encarnação, and Pinheiro (2019) looked to describe how applicants —and therefore students— to higher education perceive the proximity between degree programs using data from application processes of Portugal and Chile over several years. The authors take advantage of the nationwide application processes in both Chile and Portugal, where every year candidates submit a ranked list of degree programs and schools in which they wish to study. Applications are submitted to a centralized nationwide body that then carries the matching between applicants' preferences and schools' open positions. Hence, relying on the applicants' preferences, they propose the Higher Education Space (HES) as a means of describing a mapping of similarities between degree programs available in the Higher Education systems of Portugal and Chile. The emerging structure, the HES, is a network that connects pairs of degree programs according to the likelihood that they co-occur in the applicant's preferences. Therefore, the HES represents 'how students, not administrators or faculty, think about the grouping of' degree programs, and it does so by leveraging existing Big Data sources that are naturally produced in many systems that, such as Chile and Portugal, rely on centralized and nationwide application processes.

Figure 3 depicts the Higher Education Space along with some of the stylized facts presented by the authors in their original work (Candia, Encarnação, and Pinheiro 2019). In particular, the authors show that the resulting network structure would reveal positive assortments of characteristics of enrolled students (gender and application scores), contextual information such as reported unemployment levels, educational experience through dropout rates, but also variations in demand-supply ratio. In other words, the network showed that students with similar characteristics tended to select degree programs closer in the HES, but also that degree programs closer in the HES shared similar educational and socio-economic outcomes and experienced similar variations in demand.

The HES (Candia, Encarnação, and Pinheiro 2019) is a data-driven mapping that offers a framework that can be used not only to better understand higher education organizational features from the applicants' perspectives, but it can also be the basis for the design of natural experiments in Education Sciences and as an instrument for the improvement of the effectiveness of policy-making in higher education.

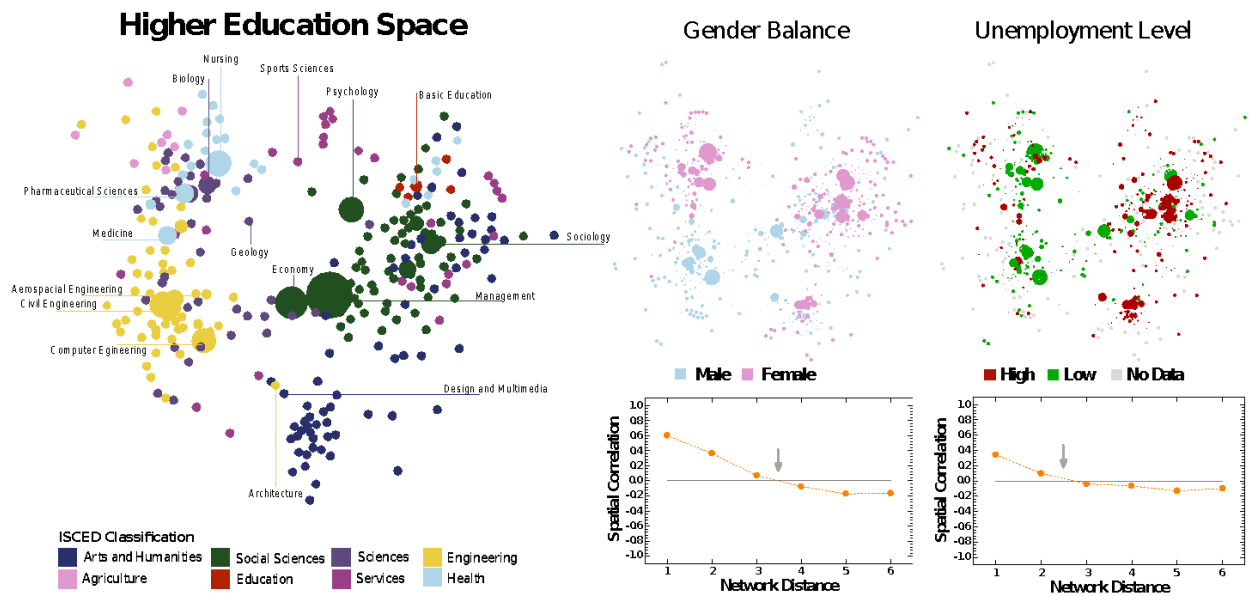

*Figure 3: The Higher Education Space (HES) is a network that connects degree programs that often appear together in the preferences of applicants to higher education. On the left, the Higher Education Space of Portugal where each node is a degree program and the edges are the existence of a significant co-occurrence in the applicants' preferences. Nodes are colored according to the ISCED classification of educational systems (Statistics 2012). On the top right, are shown the positive assortment of gender dominance among enrolled students and unemployment levels, which lead to network correlations (bottom-right) up to two or three degrees of separation.*

Such understanding of the nature of the relatedness between degree programs, schools, or careers are a decisive step into aiding school officials, applicants, and students into improving the efficiency of their choice and in evaluating the potential of future policies. In a recent project funded by FONDEF (Chile) and in collaboration with the Chilean Ministry of Education, the platform lixandria: a map for discovering your degree program, builds on top of the HES and on historical records of 1.9 million applicants to provide students with recommendations on institutions and degree programs to apply based on their preferences.

As captured in the HES, the relatedness between degree programs does not necessarily mirror formal disciplinary taxonomies defined for instance by the International Standard Classification of Education (ISCED) (Schneider 2013). The subjective interpretation of information on degree programs, along with societal pressures in the pursuit of certain professional pathways add value to [lixandria](#) as an informative tool for school interventions. For instance, school educators could expand the horizon of degree programs by engaging in early explorations of the HES and gather information on the dynamics of students' preferences throughout various years during secondary education. Data onto students' professional proclivities and motivations would not only foster more personally-sound interventions in the classroom, according to implementation science (Kelly 2012; Moir 2018; Soicher, Becker-Blease, and Bostwick 2020), but also, might boost higher levels of satisfaction and integration given the transparency of such preferences, and the further likelihood develop similarity-driven social ties (Porter and Umbach 2006; Lazarsfeld and Merton 1954; McPherson, Smith-Lovin, and Cook 2001; Smirnov, and Thurner 2017). Besides, such intervention would stimulate exploration beyond the aforementioned biases and societal pressures regarding particular groups of degree programs. In simple terms, the technology embedded in the HES affords educators with a chance to showcase both the well and less-known degree programs, and in some cases how surprisingly close and similar these are perceived in the eyes of younger generations of applicants.

Ongoing research is extending the application of the HES as a framework for academic studies as well. In that sense, recent results show that the HES can be used to quantify the degree of coherence in the applicants' preferences, which is found to be a predictor of first year dropout rates and academic retention (in-review paper), results that have been shown to be consistent both in Chile and Portugal. Moreover, the HES can be used to compare how gender prevalence on STEM enrollment has been evolving among regions, which when combined with the variability of regional socio-economic and cultural characteristics can shed light on which underlying societal factors might be more character progressing between different regions conductive of the widely reported inequalities in Engineering. First steps are being taken using the Chilean context.

In that sense, the HES shows the potential of using network analysis to extract new value and an understanding from data on how enrolled and prospective students perceive the organization of higher education. By leveraging big data of such nature, we are showing the ability to achieve results that are of both academic and societal importance.

## Network measures

The relatedness of educational components goes beyond the exclusive exploration of subjects (i.e., students, teachers and administrators), like in the HES (Candia, Encarnação, and Pinheiro 2019). From network science it is possible then to conceptualize relationships among relevant entities and forecast their influence, and the extent of such influence, within the system. Here, the use of various network measures provide critical information for interpreting such effects. Next, we introduce some of these network variables, their applicability depending on the nature of the network, and examples of interpretations relevant in education (Table 1). We define a weighted adjacency matrix for each classroom as $w_{ij}$, where each entry represents the absence of relationship ($w_{ij} = 0$) or the strength of the directed relationship ($w_{ij} \geq 1$) from student $i$ to student $j$.

*Table 1: Network metrics and their definitions. $w_{ij}$ is the number of links (direct social relationships, number of sent tokens, or significant relationships between degree programs) between students i and j, and N is the total number of students. d represents a damping factor ($d = 0.85$ following (Page et al. 1999)), and N is the number of students in each classroom.*

| Notation | Definition |
|---|---|
| $G(V_i, E)$ | G represents a network. V represents the vertex (students, i) and E is the edges. |
| $k(V_i) = \frac{1}{N} \sum_{j \neq i} w_{ji}$ | Node degree is a centrality metric that quantifies the number of connections that a student has in the forum co-exposure network. |
| $r_i = \frac{1}{N} \sum_{j \neq i} w_{ji}$ | Average in-degree (or average received cooperation in the context of experimental game theory) measures the average cooperation received by the ego ($i$). |
| $s_i = \frac{1}{N} \sum_{j \neq i} w_{ij}$ | Average out-degree (or average sent cooperation in the context of experimental game theory) measures the average sent cooperation. |
| $R_i = \frac{1}{N} \sum_{j \neq i} min[w_{ji}, w_{ij}]$ | Reciprocated weight measures the average level of reciprocity for each ego ($i$). |
| $P(V_i) = \frac{1-d}{N} + d \sum_{j=1}^{n} \frac{w_{ij} P_j}{\sum_{k=1}^{n} w_{kj}}$ | PageRank is a centrality metric that quantifies the number of connections and the importance of those connections for each student in the network. For instance, in a cooperation network, the PageRank can be used to proxy the relative social status of each student in a classroom. d represents a damping factor. The inventors of Page-Rank (Page et al. 1999) recommend setting it as $d = 0.85$. |
| $C_i = \frac{|\{e_{jk} : v_j, v_k \in N_i, e_{jk} \in E\}|}{k_i(k_i - 1)}$ | The clustering coefficient is defined as the proportion of the number of links between the neighbors of vertex $i$ divided by the |

| Notation | Definition |
|---|---|
|  | number of links that could possibly exist between them. The more neighbors are connected between them, the higher the clustering is, leading to a higher probability to access redundant information. |
| $C'_i = \sum_j w_{ij} \left[ 1 - \sum_q \frac{w_{iq}}{\sum_{j_{i \neq j}} w_{ij}} w_{jq} \right], q \neq i, j$ | Burt's constraint for an undirected network is a centrality metric that quantifies the access to structural holes in the forum co-exposure network. The higher Burt's constraint is, the higher the probability of accessing redundant information (Burt 2004). |
| $D_{P_k} = \frac{1}{N(N-1)} \sum_{i \neq j \in P_k} d_{ij}$ | Average node distance. In the context of the Higher Education Space. It measures the relatedness of degree programs in a given application list, $P_k$, by averaging the distances between all pairs of degree programs, $d_{ij}$. |

We expect that the position of the nodes in the network together with the strength of its ties, are able to modulate the capacity of the student to capture the externalities generated by their peers during the learning process (Calvó-Armengol, Patacchini, and Zenou 2009b; Henrich 2015; Candia et al. 2019; J. Pulgar, Candia, and Leonardi 2020; Javier Pulgar, Rios, and Candia 2019; Javier Pulgar et al. 2021; Traxler et al. 2020; Baldwin, Bedell, and Johnson 1997; Caprara et al. 2000; Jesper Bruun and Brewe 2013; Gašević, Zouaq, and Janzen 2013; Blansky et al. 2013; Ivaniushina and Alexandrov 2018; Stadtfeld et al. 2019b; Kassarnig et al. 2018; Berthelon et al. 2019) or, on the other hand, the consistency of their decisions for degree programs (Candia, Encarnação, and Pinheiro 2019).

The network variables presented above yield diverse interpretations. For instance, out-degree centrality as a proxy for social engagement, a process that low achieving students are more likely to engage in when compared to their high achieving counterpart (Liu, Chen, and Daian Tai 2017), whereas in the friendship network, in-degree centrality indicates popularity. Differently, PageRank in a social directed network proxies social hierarchy (in-review paper) because it integrates both the number of connections and the importance of those connections for each node of a network. In terms of creative outputs and innovations, network constraint quantifies tie redundancy for a particular node, that is, whether an actors' ties are linked to each other, thus hindering the access to new information (Burt 2004, 2005). Accordingly, higher constraint implies less opportunities to access novel information from isolated portions of the social system, which translates into a limited capability to come up with creative ideas (Javier Pulgar, Rios, and Candia 2019). Regarding the interconnectedness of the higher education system, the HES induces a natural metric of distance between degree programs, that as related nodes (i.e., degree programs) are closer in the network (Candia, Encarnação, and Pinheiro 2019). Interestingly, applicants who consider distant (or unrelated) degree programs in their applications experience a higher chance of abandoning the degree program in which they enrolled (in-review paper).

## Concluding Remarks

The dynamics and complexity of education systems calls for adding new methodological strategies for research, interventions and evaluation. Computational social sciences afford an appropriate set of tools to conduct in-depth analysis into the myriad of constituents in education and their respective relationships. Network science provides effective methods for mapping students' relationships of different kinds, and its effects over academic achievement. Beyond these individual-level applications, networks allow to explore the underlying relatedness of other educational components, such as schools, university departments, and degree programs. Therefore, the application of network science to study educational systems provides an opportunity to discover hidden structures, understand its properties and most relevant constituents (e.g., central degree programs and universities, consistency of preferences, and direct competitors between different higher education institutions), which in return fosters interventions and decision-making.

Data-driven interventions and decision-making are thought at all levels of the educational chain: at small scale like in the classroom, to schools, universities, and system-wide. Having identified the key components and structure of these networks should enable educators, administrators, and policymakers with unique information to establish alternative strategies to overcome challenges. For instance, in the dissemination and adoption of school reforms, or to increase the relevancy of less popular professional alternatives in higher education. At classroom level, information on students' social and academic hierarchies also becomes relevant for organizing students' working partners and groups, and their disposition in the classroom.

The opportunities mentioned above come with inevitable challenges on the access to data and its analysis. Whether it is at the classroom or school levels, researchers and practitioners could resort to surveys to gather students and/or teachers' networks. However, this traditional technique carries multiple issues and biases that are difficult to signal in posterior analysis, especially in younger students. Moreover, a system-wide network study entails resorting to archives and large volumes of information stored by administrators or government officials. Thus, obtaining access to such data is often shrouded in bureaucracy and institutional blockades that make it difficult for researchers to tap into the potential value it holds. These issues can be overcome by balancing the access to institutional records and the use of designs that leverages individuals' revealed preferences, such as our experimental game theory setting (see section The Game) for mapping cooperative relations, or our backbone mapping method for quantifying degree program relationships (see section The Interconnectedness of the Higher Education System).

## Acknowledgments

The authors acknowledge the financial support of FONDEF Project 19I10413 and Data Science Institute (IDS) at Universidad del Desarrollo. Also, authors acknowledge the helpful and thorough insights of Carlos Rodriguez-Sickert, Paul Leonardi, Sara Encarnaçao, Melanie Oyarzún, Joselina Davyt, Miguel Guevara, Diego Ramirez, and Abigail Umanzor.

## Further Readings

Recommended further readings:

- Candia, C., Encarnação, S. & Pinheiro, F.L. The higher education space: connecting degree programs from individuals' choices. EPJ Data Sci. 8, 39 (2019). https://doi.org/10.1140/epjds/s13688-019-0218-4

- Pulgar, J., Candia, C., & Leonardi, P. M. (2020). Social networks and academic performance in physics: Undergraduate cooperation enhances ill-structured problem elaboration and inhibits well-structured problem solving. Physical Review Physics Education Research, 16(1), 010137.

- Boda, Z., Elmer, T., Vörös, A. et al. Short-term and long-term effects of a social network intervention on friendships among university students. Sci Rep 10, 2889 (2020). https://doi.org/10.1038/s41598-020-59594-z.

- Bruun, J., & Brewe, E. (2013). Talking and learning physics: Predicting future grades from network measures and Force Concept Inventory pretest scores. Physical Review Special Topics-Physics Education Research, 9(2), 020109.

## References

Algan, Yann, Yochai Benkler, Mayo Fuster Morell, and Jerome Hergueux. 2013. "Cooperation in a peer production economy - experimental evidence from Wikipedia." Working Paper.

Arena, Michael J, and Mary Uhl-Bien. 2016. "Complexity Leadership Theory: Shifting from Human Capital to Social Capital." People and Strategy 39 (2): 22.

Arthur, W Brian et al. 1994. Increasing Returns and Path Dependence in the Economy. University of michigan Press.

Baas, Nils Andreas, and Claus Emmeche. 1997. "On Emergence and Explanation." Intellectica 25 (2): 67–83.

Bak, Per. 2013. How Nature Works: The Science of Self-Organized Criticality. Springer Science & Business Media.

Baker, Rachel. 2018. "Understanding College Students' Major Choices Using Social Network Analysis." Research in Higher Education 59 (2): 198–225.

Baldwin, Timothy T., Michael D. Bedell, and Jonathan L. Johnson. 1997. "The social fabric of a team-based M.B.A. program: Network effects on student satisfaction and performance." Academy of Management Journal. https://doi.org/10.2307/257037.

Balland, Pierre-Alexandre, Cristian Jara-Figueroa, Sergio G Petralia, Mathieu PA Steijn, David L Rigby, and César A Hidalgo. 2020. "Complex Economic Activities Concentrate in Large Cities." Nature Human Behaviour 4 (3): 248–54.

Bandura, A. 1986. Social Foundations of Thought and Action. Engelwood Cliffs, NJ: Prentice-Hall.

Bandura, A. (2001). Social cognitive theory: An agentic perspective. Annual Review of Psychology,51, 1-26.

Banister, Emma N, and Gayle J Booth. 2005. "Exploring Innovative Methodologies for Child-centric Consumer Research." Qualitative Market Research: An International Journal 8 (2): 157–75.

Barabási, A. László. 2002. Linked: The New Science of Networks. Perseus Publishing.

Barker, John, and Susie Weller. 2003. "'Is It Fun?' Developing Children Centred Research Methods." Int. J. Sociol. Soc. Policy 23 (1/2): 33–58.

Bar-Yam, Yaneer. 2002. "General Features of Complex Systems." Encyclopedia of Life Support Systems (EOLSS), UNESCO, EOLSS Publishers, Oxford, UK 1.

Beinhocker, Eric D. 2006. The Origin of Wealth: Evolution, Complexity, and the Radical Remaking of Economics. Harvard Business Press.

Benson, D., and K. Gresham. 2007. "Social Contagion Theory and Information Literacy Dissemination; a Theoretical Model." Baltimore, MA.

Berthelon, Matias, Eric Bettinger, Diana I Kruger, and Alejandro Montecinos-Pearce. 2019. "The Structure of Peers: The Impact of Peer Networks on Academic Achievement." Res. High. Educ. 60 (7): 931–59.

Biancani, Susan, and Daniel A. McFarland. 2013. "Social Networks Research in Higher Education." In Higher Education: Handbook of Theory and Research, 151—–215. London: Springer.

Blansky, Deanna, Christina Kavanaugh, Cara Boothroyd, Brianna Benson, Julie Gallagher, John Endress, and Hiroki Sayama. 2013. "Spread of Academic Success in a High School Social Network." PLoS ONE. https://doi.org/10.1371/journal.pone.0055944.

Boda, Zsófia, Timon Elmer, András Vörös, and Christoph Stadtfeld. 2020. "Short-Term and Long-Term Effects of a Social Network Intervention on Friendships Among University Students." Scientific Reports 10: 2889.

Borgatti, Stephen, Martin Everett, and Jeffrey Johnson. 2013. Analyzing Social Networks. SAGE Publications Ltd.

Borgers, Natacha, Edith de Leeuw, and Joop Hox. 2000. "Children as Respondents in Survey Research: Cognitive Development and Response Quality 1." Bulletin of Sociological Methodology/Bulletin de Méthodologie Sociologique 66 (1): 60–75.

Brewe, E., L. Kramer, and V. Sawtelle. 2012. "Investigating Student Communities with Network Analysis on Interactions in a Physics Learning Center." Physical Review Physics Education Research 8: 010101.

Bruun, J., and E. Brewer. 2013. "Talking and Learning Physics: Predicting Future Grades from Network Measures and Force Concept Inventory Pretests Scores." Physical Review Physics Education Research 9: 021109.

Bruun, Jesper, and Eric Brewe. 2013. "Talking and learning physics: Predicting future grades from network measures and Force Concept Inventory pretest scores." Physical Review Special Topics - Physics Education Research. https://doi.org/10.1103/PhysRevSTPER.9.020109.

Burgess, L. G., P. M. Riddell, A. Fancourt, and K. Murayama. 2018. "The Influence of Social Contagion Within Education: A Motivational Perspective." Mind, Brain, and Education 12 (4).

Burt, R. S. 2004. "Structural Holes and Good Ideas." American Journal of Sociology 110 (2): 349–99.

———. 2005. "The Social Capital of Structural Holes." In The New Economic Sociology, edited by M. F Guillen, R Collin, P. England, and M. Meyer, 148–92. New York: Russell Sage.

Calvó-Armengol, Antoni, Eleonora Patacchini, and Yves Zenou. 2009a. "Peer effects and social networks in education." Review of Economic Studies. https://doi.org/10.1111/j.1467-937X.2009.00550.x.

———. 2009b. "Peer Effects and Social Networks in Education." The Review of Economic Studies 76 (4): 1239–67.

Camerer, Colin F. 2003. "Behavioural studies of strategic thinking in games." Trends in Cognitive Sciences 7 (5): 225–31. https://doi.org/10.1016/S1364-6613(03)00094-9.

Camerer, Colin F, and Robin M Hogarth. 1999. "The Effects of Financial Incentives in Experiments: A Review and Capital-Labor-Production Framework." Journal of Risk and Uncertainty 19 (1-3): 7–42.

Candia, Cristian, Sara Encarnação, and Flávio L Pinheiro. 2019. "The Higher Education Space: Connecting Degree Programs from Individuals' Choices." EPJ Data Science 8 (1): 39.

Candia, Cristian, Víctor Landaeta-Torres, César A Hidalgo, and Carlos Rodriguez-Sickert. 2019. "Strategic Reciprocity Improves Academic Performance in Public Elementary School Children." arXiv Preprint arXiv:1909.11713.

Caprara, Gian Vittorio, Claudio Barbaranelli, Concetta Pastorelli, Albert Bandura, and Philip G. Zimbardo. 2000. "Prosocial foundations of children's academic achievement." Psychological Science. https://doi.org/10.1111/1467-9280.00260.

Carolan, B. V. 2014. Social Network Analysis and Education: Theory, Methods & Applications. SAGE Publications Inc.

Chaturapruek, Sorathan, Tobias Dalberg, Marissa E Thompson, Sonia Giebel, Monique H Harrison, Ramesh Johari, Mitchell L Stevens, and Rene F Kizilcec. 2021. "Studying Undergraduate Course Consideration at Scale." AERA Open 7: 2332858421991148.

Claessens, Amy, and Mimi Engel. 2013. "How Important Is Where You Start ? Early Mathematics Knowledge and Later School Success." Teachers College Record. https://doi.org/http://dx.doi.org/10.1016/j.jcin.2012.10.019.

Coie, J. D., K. A. Dodge, and J. B. Kupersmidt. 1990. Peer group behavior and social status. Cambridge University Press.

Commeford, Kelley, Eric Brewe, and Adrienne Traxler. 2021. "Characterizing Active Learning Environments in Physics Using Network Analysis and Classroom Observations." Physical Review Physics Education Research 17: 020136.

Csibra, Gergely, and György Gergely. 2011. "Natural Pedagogy as Evolutionary Adaptation." Philosophical Transactions of the Royal Society B: Biological Sciences 366 (1567): 1149–57.

Davies, D., D. Jindal-Snape, C. Collier, R. Digby, P. Hay, and A. Howe. 2013. "Creative Learning Environments in Education-a Systematic Literature Review." Thinking Skills and Creativity 8: 80–91.

Davies, Peter. 2018. Paying for Education: Debating the Price of Progress.

DeLay, Dawn, Linlin Zhang, Laura D. Hanish, Cindy F. Miller, Richard A. Fabes, Carol Lynn Martin, Karen P. Kochel, and Kimberly A. Updegraff. 2016. "Peer Influence on Academic Performance: A Social Network Analysis of Social-Emotional Intervention Effects." Prevention Science 17: 903–13.

Dickens, William T., Isabel Sawhill, and Jeffrey Tebbs. 2006. "The effecgs of investing in ealry education on economic growth." Washington, DC: The Brookings Institution.

Eagle, N., A. Pentland, and D. Lazer. 2009. "Inferring Friendship Network Structure Using Mobile Phone Data." Proceedings of the National Academy of Science 106: 15727–278.

Eckles, James E., and Eric G. Stradley. 2012. "A Social Network Analysis of Student Retention Using Archival Data." Social Psychology of Education 15: 165–80.

Fehr, Ernst, and Andreas Leibbrandt. 2011. "A field study on cooperativeness and impatience in the Tragedy of the Commons." Journal of Public Economics. https://doi.org/10.1016/j.jpubeco.2011.05.013.

Fleer, Marilyn, and Avis Ridgway, eds. 2014. Visual Methodologies and Digital Tools for Researching with Young Children: Transforming Visuality. Springer, Cham.

Freedman, J. L., and Perlick D. 1978. "Crowding, Contagion, and Laughter." Journal of Experimental Social Psychology 15 (3).

Froehlich, D. E. 2020. "Mapping Mixed Methods Approaches to Social Network Analysis in Learning and Education." In Mixed Methods Social Network Analysis: Theories and Methodologies in Learnng and Education, edited by D. E. Froehlich, M. Rehm, and B. C. Rienties, 13–24. London: Routledge.

Gašević, Dragan, Amal Zouaq, and Robert Janzen. 2013. "Choose Your Classmates, Your GPA Is at Stake!: The Association of Cross-Class Social Ties and Academic Performance." American Behavioral Scientist. https://doi.org/10.1177/0002764213479362.

Gelcich, Stefan, Ricardo Guzman, Carlos Rodríguez-Sickert, Juan Carlos Castilla, and Juan Camilo Cárdenas. 2013. "Exploring external validity of common pool resource experiments: Insights from artisanal benthic fisheries in Chile." Ecology and Society. https://doi.org/10.5751/ES-05598-180302.

Gil-White, Francisco J., and Joe Henrich. 2001. "The Evolution of Prestige." Evolution and Human Behavior. https://doi.org/10.1007/BF00992157.

Goeree, Jacob K, Arno Riedl, and Aljaž Ule. 2009. "In Search of Stars: Network Formation Among Heterogeneous Agents." Games and Economic Behavior 67 (2): 445–66.

Goertzen, R. M., E. Brewe, and L. Kramer. 2013. "Expanding Markers of Success in Introductory University Physics." International Journal of Science Education 35 (2): 262–88.

Grunspan, D. Z., B. L. Wiggins, and S. M. Goodreau. 2014. "Understanding Classrooms Through Social Network Analysis: A Primer for Social Network Analysis in Educational Research." CBE-Life Sciences Education 13: 167–78.

Henrich, Joseph. 2015. The Secret of Our Success How Culture Is Driving Human Evolution, Domesticating Our Species, and Making Us Smarter. Princeton University Press.

Hopfensitz, Astrid, and Josepa Miquel-Florensa. 2017. "Mill ownership and farmer's cooperative behavior: The case of Costa Rica coffee farmers." https://doi.org/10.1017/S1744137416000527.

Ivaniushina, Valeria, and Daniel Alexandrov. 2018. "Anti-school attitudes, school culture and friendship networks." British Journal of Sociology of Education 39 (5): 698–716. https://doi.org/10.1080/01425692.2017.1402674.

Jacobson, Michael J, James A Levin, and Manu Kapur. 2019. "Education as a Complex System: Conceptual and Methodological Implications." Educational Researcher 48 (2): 112–19.

Johnson, David W, and Roger T Johnson. 1987. Learning Together and Alone: Cooperative, Competitive, and Individualistic Learning. Prentice-Hall, Inc.

Jones, Martin H., and Toby J. Cooke. 2021. "Social Status and Wanting Popularity: Different Relationships with Academic Motivation and Achievement." Social Psychology of Education 24: 1281–1303.

Karlan, Dean S. 2005. "Using experimental economics to measure social capital and predict financial decisions." https://doi.org/10.1257/000282805775014407.

Kassarnig, Valentin, Enys Mones, Andreas Bjerre-Nielsen, Piotr Sapiezynski, David Dreyer Lassen, and Sune Lehmann. 2018. "Academic performance and behavioral patterns." EPJ Data Science 7 (1): 10. https://doi.org/10.1140/epjds/s13688-018-0138-8.

Kelly, B. 2012. "Implementation Science for Psychology in Education." In Handbook of Implementation Science for Psychology in Education, edited by B Kelly and D. Perkins, 3–12. Cambridge, UK.: Cambridge University Press.

Kyritsi, Krystallia. 2019. "Doing Research with Children: Making Choices on Ethics and Methodology That Encourage Children's Participation." J. Child. Stud., July, 39–50.

Lane, David, Denise Pumain, Sander Ernst van der Leeuw, and Geoffrey West. 2009. Complexity Perspectives in Innovation and Social Change. Vol. 7. Springer Science & Business Media.

Laninga-Wijnen, Lydia, Allison M. Ryan, Zeena Harakeh, Huiyoung Shin, and Wilma A. Volleberg. 2018. "The Moderating Role of Popular Peers; Achievement Goals in 5th- and 6th-Graders' Achievement-Related Friendships: A Social Network Analysis." Journal of Educational Psychology 110 (2): 289–307.

Lansing, J Stephen. 2003. "Complex Adaptive Systems." Annual Review of Anthropology 32 (1): 183–204.

Lazarsfeld, P. F., and R. K. Merton. 1954. "Friendship as a Social Process: A Substantive and Methodological Analysis." In Freedom and Control in Modern Society, edited by M. Berger, T. Abel, and C. H. Page. New York: Van Nostrand.

Lee, Moosung. 2010. "Researching Social Capital in Education: Some Conceptual Considerations Relating to the Contribution of Network Analysis." Briths Journal of Sociology of Education 31 (6). https://doi.org/https://doi.org/10.1080/01425692.2010.515111.

Leeuw, Edith D de. 2011. "Improving Data Quality When Surveying Children and Adolescents: Cognitive and Social Development and Its Role in Questionnaire Construction and Pretesting." In Report Prepared for the Annual Meeting of the Academy of Finland: Research Programs Public Health Challenges and Health and Welfare of Children and Young People, 10–12.

Lemke, Jay L, and Nora H Sabelli. 2008. "Complex Systems and Educational Change: Towards a New Research Agenda." Educational Philosophy and Theory 40 (1): 118–29.

Liu, C.-C., Y.-C. Chen, and S.-J. Daian Tai. 2017. "A Social Network Analysis on Elementary Student Engagement in the Networked Creation Community." Computers & Education 115: 114–25.

López Solé, S., M. Civís Zaragoza, and J. Díaz-Gibson. 2018. "Improving Interaction in Teacher Training Programmes: The Rise of the Social Dimension in Pre-Service Teacher Education." Teachers and Teaching 24 (6): 644–58.

Mason, Mark. 2008. "What Is Complexity Theory and What Are Its Implications for Educational Change?" Educational Philosophy and Theory 40 (1): 35–49.

McCormick, Meghan P, and Elise Cappella. 2015. "Conceptualizing academic norms in middle school: A social network perspective." The Journal of Early Adolescence. https://doi.org/10.1177/0272431614535093.

McPherson, M., L. Smith-Lovin, and J. M. Cook. 2001. "Birds of a Feather: Homophily in Social Networks." Annual Review of Sociology 27 (1).

Moir, Taryn. 2018. "Why Is Implementation Science Important for Intervention Design and Evaluation Within Educational Settings?" Frontiers in Education 3. https://doi.org/10.3389/feduc.2018.00061.

Moleenaar, Nienke M., and P. J. C. Sleegers. 2010. "Social Networks, Trust, and Innovation: The Role of Relationships in Supporting an Innovative Climate in Dutch Schools." In Social Network Theory and Educational Change, edited by A. J. Daly. Cambridge, MA.: Harvard Education Press.

Moreno, J. L. 1934. Who shall survive?: A new approach to the problem of human interrelations. Washington: Nervous; Mental Disease Publishing Co. https://doi.org/10.1037/10648-000.

Mouw, Ted. 2006. "Estimating the Causal Effect of Social Capital: A Review of Recent Research." Annual Review of Sociology. https://doi.org/10.1146/annurev.soc.32.061604.123150.

Neal, Jennifer Watling. 2007. "Why Social Networks Matter: A Structural Approach to the Study of Relational Aggression in Middle Childhood and Adolescence." Child & Youth Care Forum. https://doi.org/10.1007/s10566-007-9042-2.

OECD. 2020. PISA 2018 Results (Volume v). https://doi.org/https://doi.org/https://doi.org/10.1787/ca768d40-en.

Page, Lawrence, Sergey Brin, Rajeev Motwani, and Terry Winograd. 1999. "The PageRank Citation Ranking: Bringing Order to the Web." Stanford InfoLab.

Pentland, A. 2015. Social Physics: How Social Networks Can Make Us Smarter. Penguin Books. https://books.google.com/books?id=wBHcoAEACAAJ.

Porter, S. R., and P. D. Umbach. 2006. "College Major Choice: An Analysis of Person–Environment Fit." Research in Higher Education 47 (4): 429–49.

Pulgar, J. 2021. "Classroom Creativity and Students' Social Networks: Theoretical and Practical Implications." Thinking Skills and Creativity 42. https://doi.org/https://doi.org/10.1016/j.tsc.2021.100942.

Pulgar, Javier, Diego Ramírez, Abigail Umanzor, Cristian Candia, and Iván Sánchez. 2021. "Student Networks on Online Teaching Due to COVID-19: Academic Effects of Strong Friendship Ties and Perceived Academic Prestige in Physics and Mathematics Courses." arXiv Preprint arXiv:2109.06245.

Pulgar, Javier, Carlos Rios, and Cristian Candia. 2019. "Physics Problems and Instructional Strategies for Developing Social Networks in University Classrooms." arXiv Preprint arXiv:1904.02840.

Pulgar, J., C. Candia, and P. Leonardi. 2020. "Social Networks and Academic Performance in Physics: Undergraduate Cooperation Enhances Ill-Structured Problem Elaboration and Inhibits Well-Structured Problem Solving." Physical Review Physics Education Research 16: 010137.

Putnik, G., E. Costa, C. Alves, H. Castro, L. Varela, and V. Shah. 2016. "Analyzing the Correlation Between Social Network Analysis Measures and Performance of Students in Social Network-Based Engineering Education." International Journal of Technology and Design Education 26: 413–37.

Restuccia, Diego, and Carlos Urrutia. 2004. "Intergenerational persistence of earnings: The role of early and college education." American Economic Review. https://doi.org/10.1257/0002828043052213.

Rhee, L., and P. Leonardi. 2018. "Which Pathways to Good Ideas? An Intention-Based View of Innovation in Social Networks." Strategic Management Journal 39: 1188–1215.

Rizzuto, Tracey E, Jared LeDoux, and John Paul Hatala. 2009. "It's Not Just What You Know, It's Who You Know: Testing a Model of the Relative Importance of Social Networks to Academic Performance." Social Psychology of Education 12 (2): 175–89.

Roger, T, and David W Johnson. 1994. "An Overview of Cooperative Learning." Creativity and Collaborative Learning, 1–21.

Sacerdote, Bruce. 2011. Peer Effects in Education: How might they work, how big are they and how much do we know Thus Far? https://doi.org/10.1016/B978-0-444-53429-3.00004-1.

Schneider, Silke L. 2013. "The International Standard Classification of Education 2011." In Class and Stratification Analysis. Emerald Group Publishing Limited.

Smirnov, I., & Thurner, S. (2017, aug). Formation of homophily in academic performance: Students change their friends rather than performance.PLOS ONE,12(8). Retrievedfromhttp://dx.plos.org/10.1371/journal.pone.0183473doi:10.1371/journal.pone.0183473

Soicher, R. N., K. A. Becker-Blease, and K. C. P. Bostwick. 2020. "Adapting Implementation Science for Higher Education Research: The Systematic Study of Implementing Evidence-Based Practices in College Classrooms." Cognitive Research: Principles and Implications 5 (54). https://doi.org/https://doi.org/10.1186/s41235-020-00255-0.

Stadtfeld, Christoph, András Vörös, Timon Elmer, Zsófia Boda, and Isabel J Raabe. 2019a. "Integration in Emerging Social Networks Explains Academic Failure and Success." Proceedings of the National Academy of Sciences 116 (3): 792–97.

———. 2019b. "Integration in emerging social networks explains academic failure and success." Proceedings of the National Academy of Sciences of the United States of America 116 (3): 792–97. https://doi.org/10.1073/pnas.1811388115.

Stadtfeld, Christoph, András Vörös, Timon Elmer, Zsófia Boda, and Isabel J. Raabe. 2019c. "Integration in Emerging Social Networks Explains Academic Failure and Success." Proceedings of the National Academy of Sciences 116 (3): 792–97. https://doi.org/10.1073/pnas.1811388115.

Statistics, UNESCO Institute for. 2012. "International Standard Classification of Education: ISCED 2011." Comparative Social Research 30.

Sundstrom, Meagan, David G. Wu, Cole Walsh, Ashley B. Heim, and N. G. Holmes. 2022. "Examining the Effects of Lab Instruction and Gender Composition on Intergroup Interaction Networks in Introductory Physics Labs." Phys. Rev. Phys. Educ. Res. 18 (January): 010102. https://doi.org/10.1103/PhysRevPhysEducRes.18.010102.

Traxler, A. T., T. Suda, E. Brewe, and K. Commeford. 2020. "Network Positions in Active Learning Environments in Physics." Physical Review Physics Education Research 16: 020129.

Valente, T W. 2012. "Network Interventions." Science 337 (6090): 49–53.

Van de Mortel, Thea F et al. 2008. "Faking It: Social Desirability Response Bias in Self-Report Research." Australian Journal of Advanced Nursing, The 25 (4): 40.

Vignery, Kristel, and Wim Laurier. 2020. "Achievement in Student Peer Networks: A Study of the Selection Process, Peer Effects and Student Centrality." International Journal of Educational Research 99: 101499.

Vygotsky, L. S. 1978. Mind in Society: The Development of Higher Psychological Processes. Cambridge, MA: Harvard University Press.

Wang, Zhen, Marko Jusup, Rui Wu Wang, Lei Shi, Yoh Iwasa, Yamir Moreno, and Jürgen Kurths. 2017. "Onymity promotes cooperation in social dilemma experiments." Science Advances. https://doi.org/10.1126/sciadv.1601444.

Zandena, P. J. A. C. van der, P. C. Meijera, and R. A. Beghetto. 2020. "A Review Study about Creativity in Adolescence: Where Is the Social Context?" Thinking Skills and Creativity 38: 100702.

Zwolak, J. P., M. Zwolak, and E. Brewe. 2018. "Educational Commitment and Social Networking: The Power of Informal Networks." Physical Review Physics Education Research 14: 010131.